\documentclass{article}

\usepackage{arxiv}

\usepackage[utf8]{inputenc} 
\usepackage[T1]{fontenc}    
\usepackage{hyperref}       
\usepackage{url}            
\usepackage{booktabs}       
\usepackage{amsfonts}       
\usepackage{nicefrac}       
\usepackage{microtype}      
\usepackage{lipsum}
\usepackage{graphicx}
\graphicspath{ {./images/} }

\title{Max-Cut Clustering Utilizing Warm-Start QAOA and IBM Runtime}

\author{
 Daniel Beaulieu \\
  Deloitte Consulting LLP\\
  Arlington, VA 22209 \\
  \texttt{dabeaulieu@deloitte.com} \\
   \And
 Anh Pham \\
  Deloitte Consulting LLP\\
  Atlanta, GA 30303 \\
  \texttt{anhdpham@deloitte.com} \\
  \And
}

\begin{document}
\maketitle
\begin{abstract}
Quantum optimization algorithms can be used to recreate unsupervised learning clustering of data by mapping the problem to a graph optimization problem and finding the minimum energy for a MaxCut problem formulation. This research tests the "Warm Start" variant of Quantum Approximate Optimization Algorithm (QAOA) versus the standard implementation of QAOA for unstructured clustering problems. The performance for IBM's new Qiskit Runtime API for speeding up optimization algorithms is also tested in terms of speed up and relative performance compared to the standard implementation of optimisation algorithms. Warm-start QAOA performs better than any other optimization algorithm, though standard QAOA runs the fastest. This research also used a non-convex optimizer to relax the quadratic program for the Warm-start QAOA. 
\end{abstract}
\section{Introduction and Motivation}
Unsupervised learning is an important methodology used by data scientists to find emergent properties of data without requiring labelled data or specifying which variable in the data set to make predictions based on. Clustering is a important unsupervised methodology for understanding scientific and consumer data. Clustering helps with exploratory data analysis, reducing the dimensionality of data, and even segmenting customers into groups without knowledge of their consumer habits. Quantum optimization algorithms can be used to perform unsupervised clustering tasks, and cluster data in a manner similar to regularly used classical clustering algorithms like K-means. 

The motivation for this project was to test the ability for Warm Start QAOA to perform unsupervised clustering using a tasks on real world data, as opposed to randomly generated graph data points. We tested normal QAOA, VQE, Warm Start QAOA and the Qiskit Runtime API implementations of QAOA, VQE, and Warm Start QAOA algorithms. All results were bench-marked against the classical Numpy Minimum Eigensolver to provide a reference point for the performance of the quantum algorithms. The tests were run both on the IBM QASM Simulator as well as on actual quantum hardware.

\section{Innovation}
The innovation in this project is that no one has yet applies the Warm-start variant of QAOA to unsupervised learning and tests the utility of this approach. Also, VQE has not been utilized as a method for performing unsupervised clustering before on a data formulated as a graph optimization problem, and this work determines the applicability of VQE for the unsupervised clustering task.
\paragraph{}
This project also tests the usage of the Warm-start QAOA technique with a non-convex formulation of Quadratic Unconstrained Binary Optimization (QUBO) quadratic program which our data yielded. The approach in the original paper using Warm-start QAOA requires convex formulation to optimize the semidefinite QUBO quadratic program using CPLEX and use the optimization output to produce consistently better QAOA results. The approach described in this paper can optimize a non-convex semidefinite QUBO quadratic program using the GuRoBi optimizer and produces consistent and better results than traditional QAOA. 

\section{Applicability}
This research demonstrates that Warm-start QAOA can consistently perform unsupervised clustering tasks as accurately as classical algorithms. Normal QAOA and VQE were shown to be more inconsistent in their results in our data, and produced incorrect clustering results. Warm-start QAOA gains a speed improvement with run with Qiskit Runtime that offsets the greater time it takes to run Warm-start QAOA.   

With the current state of Noisy Intermediate Scale Quantum (NISQ) technology, the current approach to using quantum optimization algorithms to perform unsupervised clustering is difficult to scale to larger data sets. The current implementation of this problem with five rows of data from the Motor Trends Car data set creates a 5x5 matrix. Each additional row expands this matrix by n+1, and the matrix by n*n. This means a 10 rows would be have 100 values compared to the 25 values in a 5x5 matrix. This rapid increase in problem size means that we had to use a small data set to keep the size of the weight matrix required to perform the graph optimization problem to a size compatible with the modest quantum hardware we had available. The quantum hardware this research is based on used a seven qubit IBM quantum computer, requiring the problem to be smaller than any real-world unsupervised clustering problem. As the quantum volume of gate-based quantum computers grows, larger data sets can be utilized for this quantum optimization based unsupervised learning task. However, QAOA and VQE do not scale well, since the number of qubits required is limited by the complexity of the quantum circuits and complexity of the optimization problem \cite{FedorovVQE}.

\paragraph{}
The current work finds that quantum optimization based unsupervised clustering performs as well as classical methods, though it does not perform better than classical methods. It is theorized that with the advent of quantum computers with larger quantum volume QAOA can out perform classical algorithms by reaching an optimal solution in fewer iterations\cite{LarkinQAOA} \cite{GuerreschiQAOA}\cite{MedvidovicQAOA}). Currently there is no clear current timeline for when quantum optimization algorithms might produce better results than classical algorithms or are able to perform tasks classical computers could not perform. Quantum optimization algorithms in the future could be able handle problems to large for classical computers to handle, including unsupervised clustering tasks.  

\section{Role of Qiskit Runtime}
We also wanted to test whether IBM Qiskit Runtime API improves the processing time of quantum optimization algorithms, and if processes performed using the Qiskit Runtime API return equivalent results as those run without the Qiskit Runtime API. The hypothesis was that Qiskit Runtime API implementations of quantum optimization algorithms performs equivalently to non-Runtime quantum optimization in terms of determine low-energy and accurate solution and perform these tasks faster on a consistent basis.

\section{Future Applications}
Future applications of quantum optimization algorithms should find better solutions to NP-complete optimization problems than classical algorithms. Quantum optimization algorithms are theorized to provide optimal solutions that classical algorithms cannot, rather than just being able to handle problems to large or complex for classical algorithms. For the current unsupervised clustering problem based on a MaxCut graph optimization, QAOA is theorized to outperform classical algorithms for MaxCut problems by reaching an optimal solution in fewer steps than classical K-means clustering\cite{OtterbachClust}\cite{KhanKMeans}. This opens up opportunities for more useful unsupervised clustering results than would be possible with classical algorithms or in much shorter periods of time.

\section{Technology Stack}
The technology stack used for this work is the IBM Quantum Lab (https://lab.quantum-computing.ibm.com/). The code was created using the python notebook interface, and all work was tested on the IBM QASM Simulator, and IBM Statevector Simulator. The authors tried to run the program on IBM Qiskit Runtime enabled quantum processors, but the even the small five row data set took a extremely long amount of time to run on the freely available IBM Quantum experience due to demand for these publicly accessible machines and long queue times for jobs submitted to them.We used the IBMQ Jakarta and IBM Lagos machines quantum computers to get our results from actual quantum hardware. 

\section{Data}
The data used for this analysis is the 1974 Motor Trend US magazine data set. The clustering objective was to cluster cars that were either sports cars or economy cars. Only five rows were selected from this data set to reduce the number of jobs that would need to be submitted to quantum back-end hardware. Even a five by five matrix required many jobs to complete on the seven qubit machines available for use by the authors and took many hours due to the queue times on IBM Q systems.

\section{Methods}
\subsection{Strategies for using Quantum Optimization for Unsupervised Learning}
The strategies for using quantum optimization algorithms are based on the approach used in Khan et al. 2019\cite{KhanKMeans}. The idea is to use quantum interference, negative rotations, and destructive interference to reach an optimal unsupervised learning state in fewer iterations than a classical algorithm. 

This research uses MaxCut rather than K-means for quantum clustering, and quantum optimization algorithms along with classical optimizer to find the optimal bit string for a weighted MaxCut instance. The approach treats a set of data points to be clustered as a MaxCut problem which is fed into the quantum optimization algorithms. The classical optimizer used was SPSA due to its robust to noise on quantum hardware \cite{NanniciniPerHybrid}. The current research evaluates QAOA and Variational Quantum Eigensolver (VQE) as methods to perform quantum optimization for unsupervised learning. For more mathematical information on QAOA please refer to Farhi et al. 2014\cite{Farhi2014QAOA} and for VQE refer to Peruzzo et al. 2013\cite{PeruzzoVQE}. 

\subsection{Warm-Start QAOA}
QAOA has lacked theoretical guarantees on its performance as well as its ability to outperform classical algorithms\cite{BravyiQA}\cite{CrooksQAOA}. The idea of Warm-start QAOA is to use a continuous-valued relaxation that is positive and semi-definite to relax a convex quadratic program and find an optimal initial starting point for QAOA\cite{EggerWSQAOA}\cite{SvatoplukSDR}. Warm-start QAOA has been shown to have a higher probability of sampling the optimal solution \cite{EggerWSQAOA}. For more details on the math underlying Warm-start QAOA, see Eggers et al. 2021\cite{EggerWSQAOA}.
\subsection{Warm-Start QAOA Process}
\begin{itemize}
\item Transform the problem data into a QUBO quadratic formulation
\item Relax the QUBO formulation and its binary variables into a semi-definite convex problem which can be solved using a classical optimizer so it becomes a multiple continuous variables with a range of 0 to 1
\item Convert problem to an Ising Hamiltonian
\item Pass the relaxed semi-definite continuous variables to form the mixer operator for the Warm-started QAOA
\item Create a mixer operator 
\end{itemize}

One modification we had to make for this research is to relax the assumption that our semi-definite quadratic program is convex. The data we used did not yield itself to creating convex quadratic programs. This meant we could not use the CPLEX optimizer to relax our semi-definite QUBO quadratic program, but instead had to use the GuRoBi optimizer which does not require convex quadratic programs. The GuRoBi optimizer created an output relaxed problem that could be used to effectively warm start QAOA. While CPLEX is found to perform better under high-dimensionality problems\cite{AnandRecipieRelax}, GuRoBi was able to produce results with the non-convex quadratic problem created from our data. 

\section{Results}
\subsection{Simulator Results}

\paragraph{}
\begin{table}[]
\caption {Quantum Optimization Unsupervised Clustering Results by Algorithm and Runtime Environment on Simulator} 
\begin{tabular}{@{}lllllllll@{}}
\toprule
Simulator          & Type    & Classical & \begin{tabular}[c]{@{}l@{}}VQE \\ (Norm)\end{tabular} & \begin{tabular}[c]{@{}l@{}}QAOA \\ (Norm)\end{tabular} & \begin{tabular}[c]{@{}l@{}}VQE \\ (Runtime)\end{tabular} & \begin{tabular}[c]{@{}l@{}}QAOA \\ (Runtime)\end{tabular} & \begin{tabular}[c]{@{}l@{}}WS-QAOA \\ (Normal)\end{tabular} & \begin{tabular}[c]{@{}l@{}}WS-QAOA \\ ( Runtime)\end{tabular} \\ \midrule
Honda   Civic      & economy & 1         & 0                                                     & 1                                                      & 0                                                        & 1                                                         & 0                                                           & 0                                                             \\
Toyota Corolla     & economy & 1         & 0                                                     & 1                                                      & 0                                                        & 1                                                         & 0                                                           & 0                                                             \\
Camaro Z28         & sport   & 0         & 1                                                     & 0                                                      & 1                                                        & 0                                                         & 1                                                           & 1                                                             \\
Pontiac Firebird   & sport   & 0         & 0                                                     & 1                                                      & 1                                                        & 0                                                         & 1                                                           & 1                                                             \\
Maserati Bora      & sport   & 0         & 1                                                     & 0                                                      & 0                                                        & 0                                                         & 1                                                           & 1                                                             \\
Energy             &         & -414.86   & -410.912                                              & -117.996                                               & -411.455                                                 & -11.640                                                   & -414.857                                                    & -414.857                                                      \\
Solution Objective &         & 1163.497  & 1159.314                                              & 1159.314                                               & 859.489                                                  & 0                                                         & 1163.497                                                    & 1163.497                                                      \\
process time:      &         & 0.014     & 4.195                                                 & 0.263                                                  & 1.606                                                    & 1.508                                                     & 0.356                                                       & 1.0119                                                        \\ \bottomrule
\end{tabular}
\end{table}

\begin{table}[]
\caption {Quantum Optimization Unsupervised Clustering Results by Algorithm and Runtime Environment on Quantum Hardware} 
\begin{tabular}{@{}lllllllll@{}}
\toprule
Simulator          & Type    & Classical & \begin{tabular}[c]{@{}l@{}}VQE \\ (Norm)\end{tabular} & \begin{tabular}[c]{@{}l@{}}QAOA \\ (Norm)\end{tabular} & \begin{tabular}[c]{@{}l@{}}VQE \\ (Runtime)\end{tabular} & \begin{tabular}[c]{@{}l@{}}QAOA \\ (Runtime)\end{tabular} & \begin{tabular}[c]{@{}l@{}}WS-QAOA \\ (Normal)\end{tabular} & \begin{tabular}[c]{@{}l@{}}WS-QAOA \\ ( Runtime)\end{tabular} \\ \midrule
Honda   Civic      & economy & 1         & 0                                                     & 1                                                      & 0                                                        & 1                                                         & 0                                                           & 0                                                             \\
Toyota Corolla     & economy & 1         & 0                                                     & 1                                                      & 0                                                        & 1                                                         & 0                                                           & 0                                                             \\
Camaro Z28         & sport   & 0         & 1                                                     & 0                                                      & 1                                                        & 0                                                         & 1                                                           & 1                                                             \\
Pontiac Firebird   & sport   & 0         & 0                                                     & 1                                                      & 1                                                        & 0                                                         & 1                                                           & 1                                                             \\
Maserati Bora      & sport   & 0         & 1                                                     & 0                                                      & 0                                                        & 0                                                         & 1                                                           & 1                                                             \\
Energy             &         & -414.86   & -114.168                                              & 13.381                                                 & -34.025                                                  & -18.476                                                   & -164.519                                                    & -290.264                                                      \\
Solution Objective &         & 1163.497  & 1159.313                                              & 859.49                                                 & 0                                                        & 0                                                         & 1163.497                                                    & 1163.497                                                      \\
process time:      &         & 0.014     & 174.453                                               & 25.031                                                 & 4.125                                                    & 3.81                                                      & 81.178                                                      & 4.35                                                          \\ \bottomrule
\end{tabular}
\end{table}

\subsubsection{Base Quantum Optimization Algorithm Results}
The results show that QAOA and VQE run in the normal manner did not produce accurate clustering results compared to the ground truth in the real data or the output of the classical optimization algorithm. Normal QAOA in particular had a much higher energy than VQE or the compared to VQE or the classical algorithm implying worse performance.

When run on the seven qubit IBM Lagos system, neither QAOA or VQE assigned all observations to the correct cluster. However, also Base VQE and QAOA did not correctly assign observations to the correct cluster on the simulator, so the task performance wasn't worse on quantum hardware. When run on quantum hardware QAOA and VQE both incorrectly clustered observationso. The solution objective was lower for both QAOA and VQE on quantum hardware than on the simulator, which can be read as a rating of lower performance on quantum hardware. The process time for base VQE and QAOA was significantly longer than on the simulator, which is an expected result.

\subsubsection{Qiskit Runtime Quantum Optimization Algorithm Results}
For the runtime versions of the algorithm, VQE produced incorrect clustering results, and QAOA produced correct results, albeit with a low energy. The Qiskit Runtime versions were not expected to yield time-savings on simulator hardware, and did not run the processes faster than the non-Runtime optimization versions of the algorithms.

The Qiskit Runtime versions of QAOA and VQE run on actual quantum hardware incorrectly more records than the non-Qiskit Runtime implementation. However, it should be noted that the base versions also did not assign all records correctly. The solution objective for both VQE and QAOA when run on Qiskit Runtime as zero, which is an unexpected result. 

\subsubsection{Warm-start QAOA Algorithm Results}
Warm-start QAOA produced accurate clustering results with low energy levels as good as the classical optimizer. The Warm-start QAOA algorithm produced the best unsupervised clustering results of all quantum optimization algorithms. The Qiskit Runtime enabled version of the Warm-start QAOA performed identically to the normal implementation of Warm-start QAOA. 
\paragraph{}
On the IBM Lagos seven qubit quantum computer, the results show that Warm-start QAOA performed significantly better than normal VQE or QAOA. Warm-start QAOA correctly clustered all observations in the data set when run on quantum hardware, and provided solution objective scores that were equivalent to the classical algorithm. It is interesting to note that Warm-start QAOA performed equivalently in terms of assigning observations to the correct cluster as well as in terms of solution objective for the Qiskit Runtime and normal implementation of Warm-start QAOA. In addition, Qiskit Runtime enabled Warm-start QAOA ran 18 times faster than the normal implementation of Warm-start QAOA, providing similar results in a fraction of the time. It should be noted that Warm-start QAOA took nearly three times as long as normal QAOA when not run using Qiskit Runtime, making this methodology significantly more costly in terms of computer time when not run using Qiskit Runtime.
\paragraph{}
\begin{figure}[h]
\caption{Figure 1. Eigenstate Vectors for Quantum Optimization Algorithms on Simulator}
\centering
\includegraphics[width=17cm]{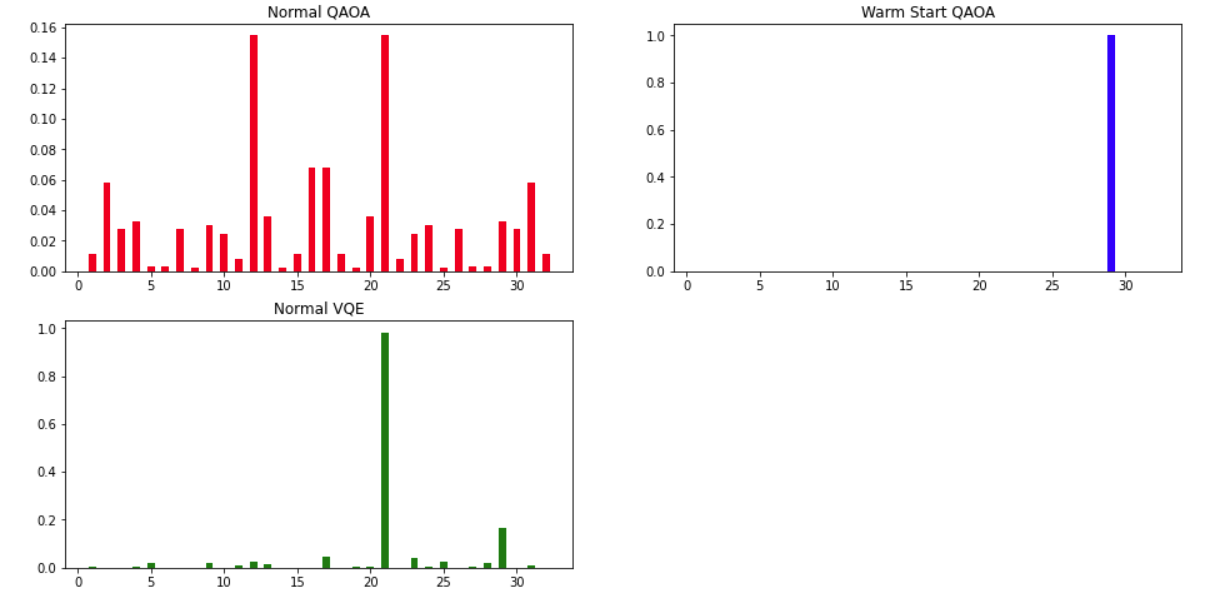}
\end{figure}
The advantage that Warm-start QAOA has over normal QAOA is demonstrated clearly in Figure 1. The normal QAOA shows multiple peaks in the Eigenstate vector energy graph which could be chosen as an optimal. The Warm-start QAOA only has one clear peak in Eigenstate vector energy, showing a clearer result. The results for Warm-start QAOA are more comparable to VQE, which also shows a pronounced Eigenstate vector energy peak. 

\section{Discussion and Conclusions}
The results of this research show that Warm-start QAOA produces more consistent results than other algorithms on quantum simulators. This work also showed that non-convex quadratic programs can be relaxed successfully using the GuRoBi optimizer and used to make QAOA consistently produce consistently better performance. 
\paragraph{}
This research also showed that Qiskit Runtime does not negatively impact the performance of quantum optimization algorithms on a quantum simulator, and in the case of Warm-start QAOA produced equivalent results 18 times faster. The equivalent performance of Warm-start QAOA running normally and through Qiskit Runtime shows Qiskit Runtime can provide equivalent performance in a fraction of the time when quantum algorithms use a Warm-start procedure. Neither VQE or normal QAOA provided correct clustering results when run on either a simulator, quantum hardware, or with Qiskit Runtime enabled. Warm-start QAOA produced correct clustering results and solution objective score equivalent to the classical algorithm regardless of whether run on the simulator, on quantum hardware, or with Qiskit Runtime enabled. Warm-start QAOA consistently showed the best performance of the quantum optimization algorithms for clustering tasks.
\paragraph{}

The results indicate that quantum algorithms can consistently produce unsupervised clustering results when using the MaxCut formation that are as good as a classical machine when using the Warm-start QAOA methodology on current low qubit NISQ systems. Qiskit Runtime is a positive step towards speeding up quantum computation. The finding that Qiskit Runtime sped up Warm-start QAOA is especially important since Warm-start QAOA took three times as long as normal QAOA to run on quantum hardware. Graph based unsupervised clustering is hypothesized to require fewer iterations to reach a optimal answer than classical computational algorithms\cite{OtterbachClust}\cite{KhanKMeans}. When the quantum computing performance and quantum volume increase, it is theoretically possible that quantum computers may be able to perform unsupervised clustering faster than classical computers. 

\bibliographystyle{unsrt}  


\paragraph{}
Legal Disclaimer:
\paragraph{}
About Deloitte
\paragraph{}
Deloitte refers to one or more of Deloitte Touche Tohmatsu Limited (“DTTL”), its global network of member firms, and their related entities (collectively, the “Deloitte organization”). DTTL (also referred to as “Deloitte Global”) and each of its member firms and related entities are legally separate and independent entities, which cannot obligate or bind each other in respect of third parties. DTTL and each DTTL member firm and related entity is liable only for its own acts and omissions, and not those of each other. DTTL does not provide services to clients. Please see www.deloitte.com/about to learn more
\paragraph{}
Deloitte is a leading global provider of audit and assurance, consulting, financial advisory, risk advisory, tax and related services. Our global network of member firms and related entities in more than 150 countries and territories (collectively, the “Deloitte organization”) serves four out of five Fortune Global 500® companies. Learn how Deloitte’s approximately 330,000 people make an impact that matters at www.deloitte.com.
\paragraph{}
About This Article
\paragraph{}
This communication contains general information only, and none of Deloitte Touche Tohmatsu Limited (“DTTL”), its global network of member firms or their related entities (collectively, the “Deloitte organization”) is, by means of this communication, rendering professional advice or services. Before making any decision or taking any action that may affect your finances or your business, you should consult a qualified professional adviser. No representations, warranties or undertakings (express or implied) are given as to the accuracy or completeness of the information in this communication, and none of DTTL, its member firms, related entities, employees or agents shall be liable or responsible for any loss or damage whatsoever arising directly or indirectly in connection with any person relying on this communication. DTTL and each of its member firms, and their
\paragraph{}
Copyright © 2021. For information contact Deloitte Global.

\end{document}